\documentclass[twocolumn,showpacs,preprintnumbers,amsmath,amssymb,prb,superscriptaddress]{revtex4-1}

\usepackage{graphicx}
\usepackage{dcolumn}
\usepackage{bm}
\usepackage{color}

\begin{document}

\title{Dispersive line shape in the vicinity of the $\nu=1$ quantum Hall state: \\Coexistence of Knight shifted and unshifted RDNMR responses}

\author{W. Desrat}
\altaffiliation{W. Desrat and B. Piot contributed equally to this work.}
\affiliation{Universit\'{e} Montpellier 2 and CNRS, Laboratoire Charles Coulomb UMR 5221, F-34095, Montpellier, France}
\affiliation{Laboratoire National des Champs Magn\'{e}tiques Intenses, CNRS-UJF-UPS-INSA, 38042 Grenoble, France}

\author{B.A. Piot}
\altaffiliation{W. Desrat and B. Piot contributed equally to this work.}
\affiliation{Laboratoire National des Champs Magn\'{e}tiques Intenses, CNRS-UJF-UPS-INSA, 38042 Grenoble, France}

\author{S. Kr\"{a}mer}
\affiliation{Laboratoire National des Champs Magn\'{e}tiques Intenses, CNRS-UJF-UPS-INSA, 38042 Grenoble, France}

\author{D.K. Maude}
\affiliation{Laboratoire National des Champs Magn\'{e}tiques Intenses, CNRS-UJF-UPS-INSA, 38042 Grenoble, France}

\author{Z.R. Wasilewski}
\affiliation{Department of Electrical and Computer Engineering University of Waterloo, Waterloo, ON, Canada}

\author{M. Henini}
\affiliation{School of Physics and Astronomy, University of Nottingham, Nottingham NG7 2RD, United Kingdom}

\author{R. Airey}
\affiliation{Department of Electronic and Electrical Engineering, University of Sheffield, Sheffield S1 4DU, United
Kingdom}

\date{\today}

\begin{abstract}
The frequency splitting between the dip and the peak of the resistively detected nuclear magnetic resonance (RDNMR)
dispersive line shape (DLS) has been measured in the quantum Hall effect regime as a function of filling factor,
carrier density and nuclear isotope. The splitting increases as the filling factor tends to $\nu=1$ and is proportional
to the hyperfine coupling, similar to the usual Knight shift versus $\nu$-dependence. The peak frequency shifts
linearly with magnetic field throughout the studied filling factor range and matches the unshifted substrate signal,
detected by classical NMR. Thus, the evolution of the splitting is entirely due to the changing Knight shift of the dip
feature. The nuclear spin relaxation time, $T_1$, is extremely long ($\sim$hours) at precisely the peak frequency.
These results are consistent with the local formation of a $\nu=2$ phase due to the existence of spin singlet $D^-$
complexes.
\end{abstract}

\maketitle

Resistively detected nuclear magnetic resonance (RDNMR) is well-established technique to probe the interaction between
nuclear and electronic spin systems via the contact hyperfine interaction. For the investigation of two-dimensional
electron gases (2DEGs), RDNMR has significantly increased sensitivity with respect to classical NMR, as it only probes
nuclei which have significant overlap with the electronic wave function. Historically, RDNMR was developed to
demonstrate the role played by nuclear spins in the formation of a huge longitudinal resistance spike at filling factor
$\nu=2/3$~\cite{Kronmuller1999} and has since proved to be a formidable tool to investigate quantum Hall
physics.~\cite{Desrat2002,Stern2004,Masubuchi2006,Kodera2006,Tracy2007,Guo2010,Zhang2007,Dean2009,Kawamura2009,Bowers2010,Yang2011,Tiemann2012,Stern2012}

In particular, attention has been focussed on the $\nu=1$ QH state following the observation of an unexpected
dispersive line shape (DLS).~\cite{Desrat2002} A strong coupling of nuclear spins to low-energy (gapless) excitations
of the many body quantum Hall ground state in the vicinity of $\nu=1$ was invoked to explain the anomalous resonance
shape. Notably, to explain the unusually short nuclear spin relaxation times $T_1$ observed, it was suggested that a
coupling occurs with the Goldstone modes of a skyrme crystal. However, in later investigations the DLS was detected at
$\nu=1$ under unfavorable conditions for the formation of skyrmions.~\cite{Bowers2010} It is now evident that skyrmions
alone cannot account for the DLS and indeed a coherent description of the origin of the dispersive line shape is still
lacking. Thermal effects have been put forward by Tracy {\it et al.} to explain the coincidence between the DLS shape
inversion and the $dR_{xx}/dT$ sign change.~\cite{Tracy2006} Separate mechanisms for the dip and peak components have
been proposed, since they show different $T_1$ behaviors versus filling factor~\cite{Kodera2006} and respond
differently to dc-current.~\cite{Dean2009} Recently, the frequency splitting between the dip and the peak of the DLS,
at a constant filling factor, was shown to increase linearly versus the magnetic field in the $5-16$ T
range.~\cite{Bowers2010}

Here we report on the systematic investigation of dispersive line shapes around $\nu=1$ as a function of filling
factor, electron density, and nuclear isotope for three different samples at mK temperatures. The frequency splitting
between the dip and peak resonances in the magnetoresistance is shown to increase as $\nu \rightarrow 1$ and is
proportional to the carrier concentration and the hyperfine interaction. More precisely, we demonstrate  the peak
frequency shifts linearly with magnetic field and coincides with the response of nuclei which are not coupled to a
polarized electronic system, obtained by classical NMR. In contrast, the dip feature shows significant deviation from
linearity reflecting the variation of the polarization of the electronic system (Knight shift). Extremely long nuclear
spin relaxation times ($T_1 \sim 1$ hr) are also measured at the peak frequency consistent with the local formation of
a $\nu=2$ (unpolarized) phase due to the existence of spin singlet $D^-$ complexes.

RDNMR has been performed on three GaAs/AlGaAs heterojunctions patterned into Hall bars. The carrier densities and mobilities are
$1.56$, $1.63$ and $2.13\times10^{11}$ cm$^{-2}$ and $5.8$, $1.46$ and $0.82\times10^6$ cm$^2$V$^{-1}$s$^{-1}$ for samples $\sharp1$ (NRC 1707), $\sharp2$(NRC V0050) and $\sharp3$
(NU 2077) respectively.\cite{note4} Each device was placed in the mixing chamber of a dilution fridge with a base temperature of
$T=10$~mK. The radio-frequency (RF) field was applied by means of a copper coil wound around the sample and connected
to an RF-synthesizer by a rigid coaxial cable. The typical RF power used was around $-5$ dBm, which gives a significant
RDNMR signal while avoiding excessive heating (the electron temperature is estimated to be $T_{e}\approx100$ mK). All
resistive measurements were performed using a lock-in amplifier with an \emph{ac} driving current of $20-50$~nA at $f
\simeq 10.7$~Hz.

\begin{figure}[t]
\includegraphics[width=0.8 \columnwidth]{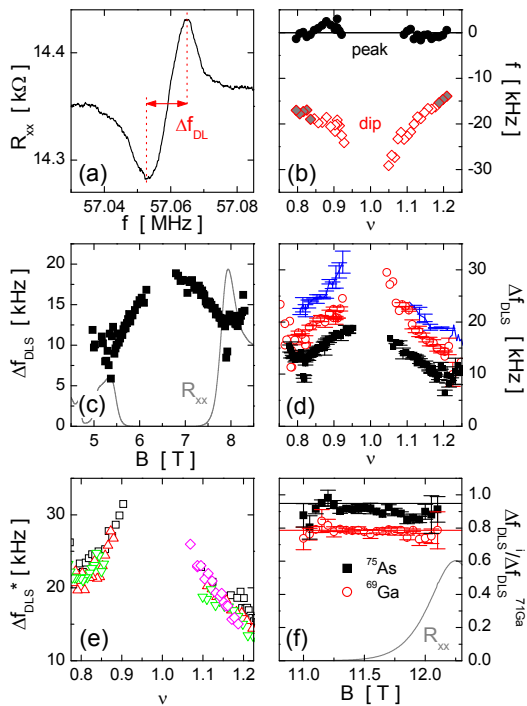}
\caption{\label{fig1} (a) Typical dispersive line shape measured by RDNMR at $\nu=0.83$. $\Delta f_{DLS}$ is the
frequency splitting between the resistance dip and peak. (b) Dip and peak frequencies of the DLS (open diamonds and
filled circles resp.) versus filling factor for $\sharp 2$. The frequency origin is arbitrary. The closed diamonds are
obtained without dc-current. (c) Frequency splitting $\Delta f_{DLS}$ as a function of the magnetic field around
$\nu=1$ for $\sharp 1$. The solid line shows $R_{xx}(B)$ plotted in arbitrary units (the horizontal axis corresponds to
$R_{xx}=0$). (d) Frequency splitting vs $\nu$ for all 3 samples ($\sharp1$, $\sharp2$, $\sharp3$ from bottom to top).
(e) $\Delta f_{DLS}\times n_s^{dark}/n_s^{light}$ vs $\nu$ for different densities (sample $\sharp 3$). (f) Frequency
splitting ratios $\Delta f_{DLS}^{75As}/\Delta f_{DLS}^{71Ga}$ and $\Delta f_{DLS}^{69Ga}/\Delta f_{DLS}^{71Ga}$
(filled squares and open circles resp.) vs $B$ on the high field flank of $\nu=1$ for $\sharp 3$. Horizontal lines
correspond to the expected ratios.}
\end{figure}

Figure \ref{fig1}(a) shows a typical RDNMR dispersive line measured for $^{75}$As at $B=7.86$~T ($\nu=0.83$) with an RF
sweep rate $df/dt=160$ Hz/s. The DLS, which is characterized by a resistance dip followed by a peak at higher radio
frequency, is observed on both sides of $\nu=1$ for all three samples. Successive RDNMR spectra have been recorded at
several magnetic fields under the same temperature and RF sweep rate conditions. The nuclear resonant frequency is
given by $f=\frac{\gamma}{2\pi}(B+B_e)$, where $\gamma$ is the nuclear gyromagnetic ratio and $B_e$ the effective
electronic field seen by the nucleus due to the contact hyperfine interaction. A close inspection reveals that the peak
response shifts linearly with magnetic field suggesting that it is not influenced by the polarization of the electronic
system. Assuming that the influence of $B_e$ is negligible for the peak response, the Larmor frequency can be
subtracted from the RDNMR data, by fitting a linear dependence to the peak resonance frequency versus $B$.~\cite{note0} The
frequency positions of the DLS peak and dip from which the Larmor frequency has been subtracted are plotted in
Fig.~\ref{fig1}(b) as a function of the filling factor for sample $\sharp2$. We observe that the peak frequency remains
constant throughout the $0.8-1.2$ filling factor range. In contrast, the dip frequency is shifted to lower frequencies
as $\nu$ tends to one \emph{from either above or below}. We should stress that the detection of the RDNMR deep into the
quantized regime (QHE regime of dissipationless magnetoresistance, see Fig. \ref{fig1}(c)) was made possible by
applying increasing dc currents (up to $5$ $\mu$A), as described by Dean {\it et al.}.~\cite{Dean2009} The closed
diamonds in Fig.~\ref{fig1}(b) represent data obtained without dc current far from $\nu=1$.

\begin{figure}[h]
\includegraphics[width=0.8\columnwidth]{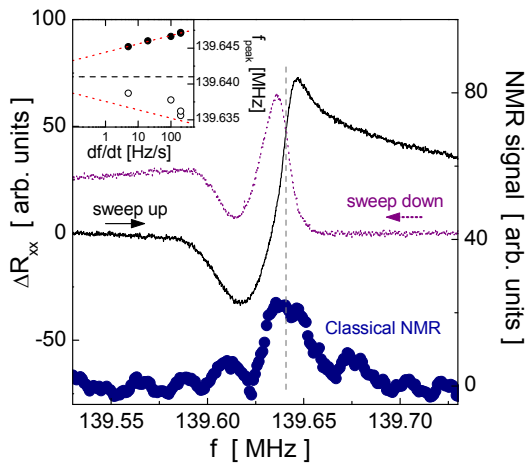}
\caption{\label{fig2} RDNMR dispersive lines (upper curves, left axis) and Fourier transform of the free induction
decay signal (filled circles, right axis) of $^{71}$Ga measured at $B=10.8$~T on sample $\sharp3$. The vertical dashed
line shows the NMR substrate frequency and arrows stand for the RF sweep directions. In inset, DLS peak frequencies for
RF-upsweeps (disks) and downsweeps (circles) vs RF sweep rate. The dashed line shows the classical NMR signal
frequency. Dotted lines are a guide to the eye.}
\end{figure}

The very different dependence of the peak and dip frequencies upon the filling factor suggests they originate from
different NMR responses.~\cite{Dean2009} In order to confirm this point, classical NMR has been performed on sample
$\sharp 1$. Figure \ref{fig2} shows the Fourier transform of the free induction decay signal of $^{71}$Ga nuclei
obtained by classical NMR and RDNMR performed at exactly the same magnetic field. The peak of the NMR signal indicates
the resonant frequency of the majority nuclei in the sample, \emph{i.e.} those in the undoped barriers and substrate,
which are not coupled to conduction electrons. In contrast, the RDNMR signal is only sensitive to the nuclei which are
in contact with the wave function of the 2DEG. These nuclei will feel the spin polarization of the 2DEG and their NMR
frequency should be Knight shifted accordingly. The peak position of the RDNMR line for the RF upsweep occurs at a
higher frequency than for the RF downsweep, but this discrepancy reduces as the RF sweep rate becomes slower. The inset
of Fig.~\ref{fig2}, which plots the DLS peak position for up and down sweeps as a function of the RF sweep rate,
suggests the peak positions will converge to the reference NMR frequency for an experimentally unrealistic sweep rate
of 10 mHz/s. Thus, the peak response originates from nuclei which are in contact with the 2DEG, but nevertheless see
zero electronic polarization ($B_e=0$) over a wide range of filling factors.


The frequency splitting of the dispersive line, $\Delta f_{DLS}$, \emph{i.e.} the absolute difference between the dip
and peak frequencies; plotted in Fig. \ref{fig1}(a); was found to be perfectly reproducible for a given constant RF
sweep rate. Figure \ref{fig1}(c) shows that $\Delta f_{DLS}$ increases when the filling factor tends to $\nu=1$ from
either above or below. In Fig. \ref{fig1}(d), $\Delta f_{DLS}$ is plotted vs $\nu$ for all three samples. They all
exhibit a similar behavior, which is reminiscent of the Knight shift plots obtained previously by optically pumped NMR
and standard NMR with multiple-quantum-well samples.~\cite{Barrett1995,Melinte2001} The Knight shift is proportional to
the electron spin polarization $\mathcal{P}$ and the hyperfine coupling which can be written
$A/h=\frac{4}{3}\mu_0g_0\mu_B\gamma\vert\Psi_0\vert^2$. In two dimensions, it has to be corrected by the factor
$n_s/w$, where $w$ is the 2DEG width. In Fig. \ref{fig1}(d), the absolute value of $\Delta f_{DLS}$ is larger for
higher density 2DEGs, which is in qualitative agreement with the above picture. A quantitative comparison cannot be
performed as the extent of the 2DEG electronic wave functions are not precisely known.

For sample $\sharp3$ we have also measured the frequency splitting as a function of the 2DEG density $n_s$, which was
modified by successive {\it in situ} illuminations with an infrared LED, using the persistent photo-conductivity
effect. Fig.~\ref{fig1}(e) shows the frequency splitting multiplied by the density ratio $n_s^{dark}/n_s^{light}$ for
densities $2.13$, $2.66$, $2.90$ and $3.26\times10^{11}$ cm$^{-2}$ (squares, up and down triangles, and diamonds
resp.). The superimposition of all the data sets confirms that as for a Knight shift, the DLS frequency splitting is
proportional to the electron density. Finally, the ratio of the frequency splitting of the $^{75}$As and $^{69}$Ga
isotopes in respect of $\Delta f_{DLS}$ for $^{71}$Ga are plotted in Fig.~\ref{fig1}(f), for the high magnetic field
flank of $\nu=1$ (sample $\sharp3$). We see that for each nuclear species the ratio is almost constant over the
measured field range, i.e. $0.899$($0.773$) for $^{75}$As($^{69}$Ga), in agreement within $5\%$ with the expected
$\gamma\vert \Psi_0\vert^2$ values ($0.948$ and $0.787$ resp.).\cite{note2} To summarize, Figs.~\ref{fig1}(d-f)
demonstrate unambiguously that the DLS frequency splitting $\Delta f_{DLS}$ depends on the filling factor $\nu$, the
electron density $n_s$ and the hyperfine interaction $A$, in \emph{exactly the same way as the Knight shift}. Note that
the measured frequency splittings $\Delta f_{DLS}$ range within $5-30$ kHz, which agree well the expected hyperfine
splittings $A$ in 2DEGs, scaled by the $n_s/w$ factor.

\begin{figure}[b]
\includegraphics[width=0.8\columnwidth]{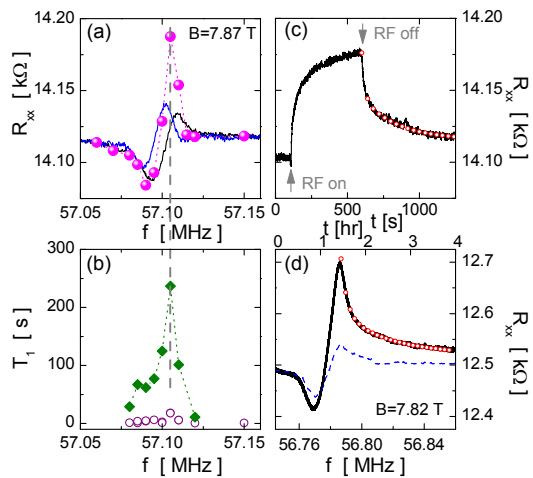}
\caption{\label{fig3}(a) DLS at $B=7.87$ T for RF up/downsweeps (solid lines) and under static conditions (dots). (b)
Slow and fast $T_1$ vs radio-frequency $f$ (filled diamonds and circles resp.). The vertical dashed line indicates the
peak resonance position at $f=57.105$ MHz. (c) Time dependence of $R_{xx}$ for an RF on/off sequence at the peak
frequency $f=57.105$ MHz. Double exponential relaxation fit (circles). (d) DLS measured at $B=7.82$ T with a slow
RF-sweep rate of $8$ Hz/s (solid line). The top axis indicates the elapsed time in hours. For comparison, the DLS
measured with a 100 times faster sweep rate ($800$ Hz/s) is plotted (dashed line). Double exponential fit of the slow
trace relaxation (circles).}
\end{figure}

The fact that the frequency position of the DLS peak remains constant versus filling factor around $\nu=1$ and matches
the classical NMR reference frequency demonstrates unequivocally that it reflects the response of nuclei which although
in contact with the 2DEG feel no nett polarization of the electronic system. To further probe this state, we have
measured the transient resistance when the RF is switched on and off, throughout the DLS resonance (see
Ref.\cite{Desrat2002} for details of this technique). The resistance transient during an RF on/off sequence is plotted
in Fig. \ref{fig3}(c). The frequency used here is $f=57.105$ MHz, which corresponds to the resonant frequency of the
DLS peak for $B=7.87$ T (Fig. \ref{fig3}(a)). When the RF is switched on, the resistance first increases rapidly and
then drifts slowly without ever reaching a steady state on an experimentally viable timescale (several hours). This
behavior is distinctive of frequencies close to the peak feature. For other frequencies and under the same experimental
conditions, the resistance always saturates. Similar observations were reported by Kodera {\it et al.}, who found long
relaxation time scales for the DLS peak.~\cite{Kodera2006}

The resistance relaxation, after the RF is switched off, is well fitted by a double exponential (symbols in Fig.
\ref{fig3}(c)). Both slow and fast nuclear spin relaxation times extracted from these fits are plotted in Fig.
\ref{fig3}(b). It is evident that the slow $T_1$ shows a sharp maximum at $f=57.105$ MHz which is exactly the DLS peak
position indicated on the figure by the vertical dashed line. The involvement of an extremely slow process is further
confirmed by the dispersive line recorded at a very low RF-sweep rate ($8$ Hz/s) plotted in Fig.~\ref{fig3}(d). The
total scan duration is 4 hours (top axis). Assuming that the return to equilibrium is mainly limited by $T_1$, the
double exponential resistance relaxation, above the peak frequency, leads to a characteristic time of the order of $1$
hr. In addition, it is interesting to note that the DLS with an RF-sweep rate of $800$ Hz/s (dashed line) has a
frequency splitting $\Delta f_{DLS}$ almost equal to the DLS recorded with a $100$ times slower sweep rate. This
justifies \emph{a posteriori} the choice of an intermediate sweep rate of $160$ Hz/s for the measurement of the
frequency splittings $\Delta f_{DLS}$ presented in Fig.~\ref{fig1}.

Thus, the dispersive line shape should be understood as the NMR responses of nuclei interacting with two electron
subsystems of the 2DEG. The dip corresponds to the Knight shifted resonance of nuclei coupled to partially spin
polarized electrons. The observed shift is peaked at filling factor $\nu=1$ for which the polarization of the 2DEG is a
maximum. The fast nuclear spin relaxation times observed imply that the electron system has low energy spin excitations
which favors nuclear spin relaxation via the flip-flop process in which energy has to be conserved. The Korringa law
predicts that $T_1^{-1}\propto D_\uparrow(E_f)D_\downarrow(E_f)$ is proportional to the density of spin up and spin
down states at the Fermi energy which precludes the presence of a significant spin gap in the electronic system. Within
a single electron picture, on both sides of filling factor $\nu=1$, the Fermi energy lies in a partially occupied spin
Landau level and an electronic spin excitation requires an energy $g\mu_B B$ which is orders of magnitude larger than
the nuclear spin flip energy. However, the physics in the vicinity of filling factor $\nu=1$ is known to be extremely
rich with the possibility to form spin reversed ground states \emph{i.e.} spin textures such as skyrmions which are
thought to have considerably smaller gaps for electronic spin flip excitations~\cite{Desrat2002}. The single electron
picture also predicts that the 2DEG should be completely polarized for filling factors below $\nu=1$. In reality many
body physics dominates and it is experimentally well established that the electronic polarization collapses rapidly on
either side of $\nu=1$.~\cite{Plochocka2009}

On the other hand, the peak feature of the DLS is the unequivocal signature of a zero Knight shifted NMR of nuclei
coupled to an unpolarized 2DEG. For the range of filling factors investigated this is simply not possible for the case
of electrons in uniformly occupied Landau levels. Thus, the 2DEG has to form domains with polarized and unpolarized
regions. In addition, these domains have to be localized on the time scale of the NMR measurements to prevent the
nuclei simply feeling the average spin polarization of the 2DEG. While it would not be surprising that the 2DEG could
spontaneously break symmetry to form a novel ground state with the presence of domains, there is actually a much
simpler explanation. Recent inelastic light scattering experiments have demonstrated the preponderant role played by
$D^-$ complexes in quantum Hall states and the existence of a depolarized electron subsystem at $\nu=1$ in
particular~\cite{Zhuravlev2008}. These complexes are formed by two electrons of opposite spins bound to a positively
charged donor impurity located in the barrier. The electrons remain in the lowest Landau level of the 2DEG and this
spin singlet state can be thought of as a local region in which the filling factor is $\nu=2$. The $D^-$ electrons are
weakly bound and extend over many nuclei with a Bohr radius $a_B\approx400$~$\AA$.~\cite{Huant1990} This should be
compared with the magnetic length $\ell_B = \sqrt{h/eB} \approx 230~\AA$ at $B=8$~T. Other electrons in the 2DEG are
repelled from $D^-$ both by the Pauli principle and the Coulomb repulsion. At $\nu=2$ the Fermi energy lies in the
cyclotron gap so that Korringa predicts a very long $T_1$ since nuclear spin relaxation can only proceed via slow
processes such as nuclear spin diffusion. We note that very recent experimental RDNMR results near filling factor
$\nu=2$ have revealed a reduced Knight shift, as well as long relaxation times and were interpreted as the formation of
an electron solid phase.~\cite{Rhone2013}

To conclude, the splitting of the peak and dip features of the dispersive line shape RDNMR signal in the vicinity of
$\nu=1$ can be used to determine the Knight shift and thus the spin polarization of the 2DEG. Classical NMR confirms
that the peak feature, which originates from nuclei in contact with an unpolarized phase of the 2DEG, provides a
convenient relative zero Knight shift reference signal.


\begin{thebibliography}{25}
\bibitem[{not({\natexlab{b}})}]{note0}
\bibinfo{note}{The external magnetic field was additionally calibrated using an
  $^{27}$Al reference placed below the sample.}
\end{thebibliography}

\end{document}